\documentclass{jpsj-suppl}
\usepackage{txfonts} 
\usepackage{graphicx}
\usepackage{xspace}
\usepackage{sidecap}

\newcommand{\Al}{$^{26}$Al\xspace}

\title{The \Al Gamma-ray Line from Massive-Star Regions}

\author{{Thomas \textsc{Siegert}$^{1}$} and {Roland \textsc{Diehl}$^{1,2}$}}

\inst{$^{1}$Max Planck Institut f\"ur extraterrestrische Physik, D-85748 Garching, Germany \\ 
$^{2}$Excellence Cluster 'Universe', D-85748 Garching, Germany }

\email{tsiegert@mpe.mpg.de}

\recdate{September 27, 2016}

\abst{The measurement of gamma rays from the diffuse afterglow of radioactivity originating in massive-star nucleosynthesis is considered a laboratory for testing models, when specific stellar groups are investigated, at known distance and with well-constrained stellar population. Regions which have been exploited for such studies include Cygnus, Carina, Orion, and Scorpius-Centaurus. The Orion region hosts the Orion OB1 association and its subgroups at about 450~pc distance. We report the detection of \Al gamma rays from this region with INTEGRAL/SPI.}

\kword{nucleosynthesis, radioactivity, massive stars}

\begin{document}
\maketitle

\section{Introduction}
Measurements of nuclear line emission of cosmic origins enable us to investigate where and how new interstellar nuclei are produced and released. \Al radioactivity  with its characteristic $\gamma$-ray line at 1808.73~keV and decay time $\tau$ of $\sim$1~Myr shows ongoing nucleosynthesis in our Galaxy, and is ideal to trace how ejecta are recycled from their nucleosynthesis sources into next-generation stars.

The \emph{CGRO} mission with the imaging Compton telescope instrument \emph{'COMPTEL'} obtained  a sky map of \Al $\gamma$-ray emission  \cite{Diehl:1993a,Pluschke:2001c}.
The Galaxy's \Al content as a whole then can be considered as probably being in a steady state, as many individual and independent sources contribute, and star formation is uncorrelated among local star forming regions  across the Galaxy. 
But already from the COMPTEL \Al image, showing a clumpy structure, it had been concluded that massive star groups are the dominating \Al producers in the current Galaxy, and may individually not be in a steady state; rather, the age of the specific stellar populations would determine the current amount of \Al in such a region \cite{Voss:2009}. 
Measurements of  systematic Doppler shifts of the line with Galactic longitude \cite{Diehl:2006d,Kretschmer:2013} had shown that the observed \Al $\gamma$ rays originate from sources throughout the Galaxy, including its distant and otherwise occulted regions at and beyond the inner spiral arms and bulge. Moreover, the Doppler shifts of the \Al-line centroid energy were found larger than expected from large-scale Galactic rotation, and suggested that large cavities around massive star groups play a major role in guiding ejecta flows from massive-star and supernova nucleosynthesis  \cite{Krause:2015,Krause:2016}.


The study of \Al from specific regions first focused on Cygnus \cite{Knodlseder:2000,Pluschke:2000,Pluschke:2001a}, which stands out as an individual source region in the COMPTEL \Al skymap. Population synthesis allowed comparison of the predicted impacts of massive star groups onto their surroundings, including nucleosynthesis ejecta, and also kinetic energy from winds and explosions as well as ionising starlight, to observations in a variety of astronomical windows and tracers of such massive-star action \cite{Voss:2009}. Detailed population synthesis and multi-wavelength studies of the Cygnus region \cite{Martin:2008,Martin:2009,Martin:2010b} have been followed by studies of Carina \cite{Voss:2012}, Orion \cite{Voss:2010a}, and Scorpius-Centaurus \cite{Diehl:2010} regions.



   \begin{SCfigure}  
   \centering
   \includegraphics[width=0.48\textwidth]{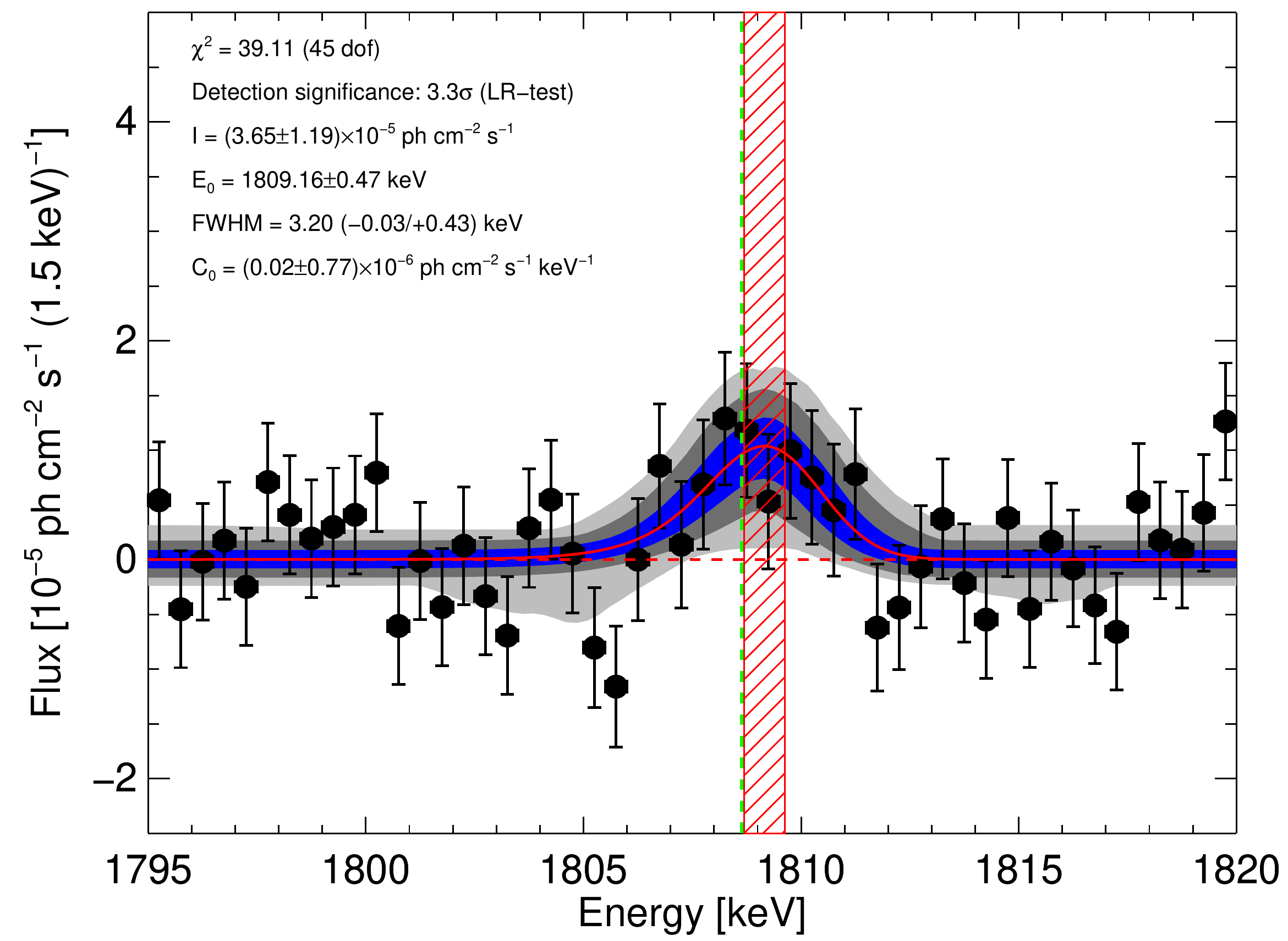}
\caption{\Al spectrum from the Orion region, as measured with SPI on INTEGRAL. Goodness of fit (through the $\chi^2$ value), detection significance, and line parameters are given in the legend. 
The blue/grey/light-grey-shaded regions show the 1$\sigma$/2$\sigma$/3$\sigma$ uncertainty range of the Gaussian fit to the line; the hatched region shows the range of blue shift that is indicated by this measurement, comparing to the laboratory value of \Al decay at rest (green dashed line). This suggests that \Al appears to stream within the Eridanus cavity from Orion OB1 stars towards the Sun.}
    \label{Fig_26Al_Orion}%
   \end{SCfigure}

\section{The Orion region}
The Orion region has been a prominent nearby region for the study of massive stars \cite{Genzel:1989}. The Orion OB1 association and its four identified subgroups  \cite{Brown:1994} originate from a parental molecular cloud now visible as Orion A and B clouds. A large interstellar cavity, called Eridanus, extends away from these molecular clouds towards the Sun, and is constrained though X-ray \cite{Burrows:1993} and HI data \cite{Brown:1995}. Our recent re-analysis of X-ray emission from this region and its comparison with 3D hydrodynamical simulations showed that energy injection and cooling of the superbubble interiors are not in a steady state \cite{Krause:2014a}.  

COMPTEL data had revealed a weak \Al $\gamma$-ray signal from this region \cite{Diehl:2003e}, attributed to nucleosynthesis ejecta from one of the Orion OB1 subgroups and its stars; the \Al $\gamma$-ray image from COMPTEL suggested that \Al was found offset from the stars of the OB1 association and probably streaming into the Eridanus cavity. 
From these multi-wavelength constraints, a population synthesis analysis was done to predict the \Al signal in more detail \cite{Voss:2010a}.
An INTEGRAL observing program was then set up, aiming at a confirmation of \Al $\gamma$ rays from Orion, with the hope of measuring kinematic information about the \Al enriched ejecta through line centroid and width determination. 

\begin{SCfigure}  
   \centering
   \includegraphics[width=0.6\textwidth]{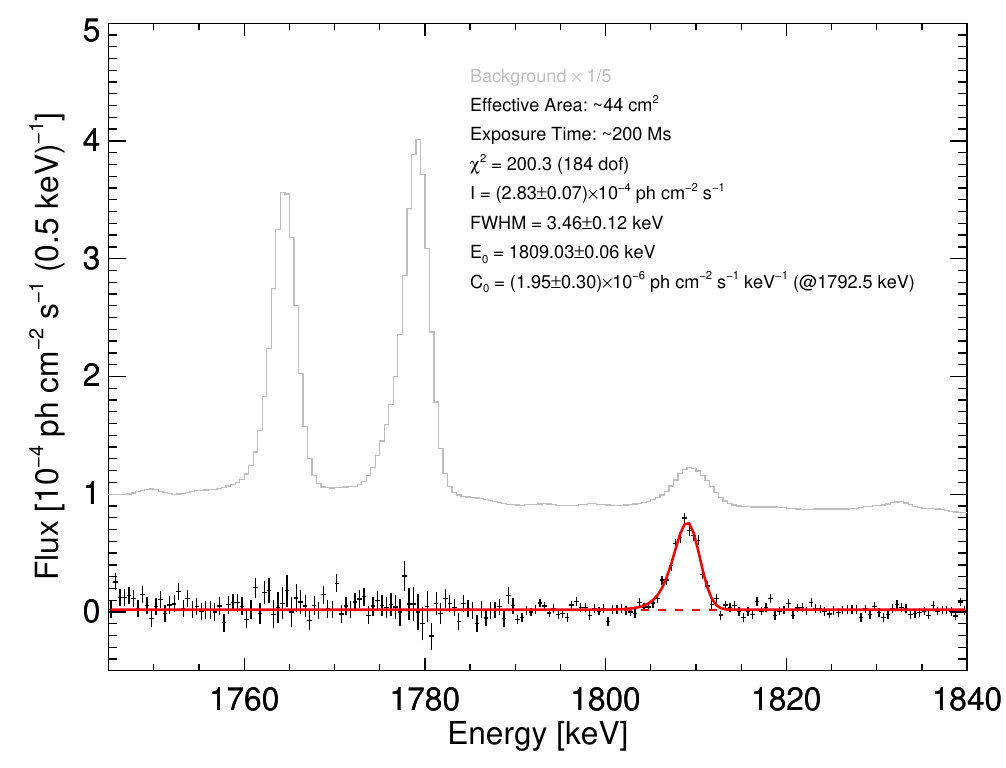}
 \caption{The INTEGRAL/SPI $\gamma$-ray spectrum near the line from \Al decay in interstellar gas  (E=1808.73~keV, $\tau$=1~Myr), compared to the spectrum of instrumental background.  The \Al emission from the sky  (red line; fitted intensities of an adopted spatial distribution model per 0.5 keV energy bin with error bars) is well discriminated against instrumental background with strong lines at 1764, 1779, and 1810 keV (grey histogram; flux values scaled by 1/5).}
   \label{Fig_26Al}%
   \end{SCfigure}

Data from 13 years of SPI single-detector events, analysed with a high spectral resolution background method, show the Galactic \Al line clearly, and demonstrate that instrumental backgrounds are well suppressed \cite{Diehl:2017} (see Fig.~\ref{Fig_26Al}).
From analysis of almost 6~Ms of data collected in 2013-2015, the COMPTEL \Al $\gamma$-ray signal from the Orion region now has been confirmed with INTEGRAL \cite{Siegert:2017} (Fig.~\ref{Fig_26Al_Orion}): the \Al line emission is detected with a significance of 3.3~$\sigma$.     

The Eridanus cavity presents a viewing geometry similar to what we propose for massive star groups in the Galaxy's spiral arms to explain the large excess above Galactic rotation velocities \cite{Kretschmer:2013,Krause:2015}. 
Fitting a Gaussian line shape to the Orion \Al $\gamma$-ray signal, the line width is found to be compatible with the instrumental resolution, and it is indicated (hatched region in Fig.~\ref{Fig_26Al_Orion})  that bulk motion may be directed towards the solar system: A corresponding blue-shift with respect to the laboratory energy value is expected from the association/cavity geometry \cite{Voss:2010a}.
 Such a blue shifted \Al signal supports the scenario we proposed for \Al ejecta streaming throughout the Galaxy, being dominated by transport within superbubbles around the massive-star groups, which extend  asymmetrically around these ejection sites, i.e., they are more elongated towards the leading edges of spiral arms and into the inter-arm regions.

\section*{Acknowledgments}
We appreciate the support from ASI, CEA, CNES, DLR, ESA, INTA, NASA and OSTC of the INTEGRAL ESA space science mission with the SPI spectrometer instrument project. 
This work was also supported from the Munich cluster of excellence \emph{Origin and Evolution of the Universe}.



\newcommand{\actaa}{Acta Astron. }%
\newcommand{\araa}{Ann.Rev.Astron.\&Astroph. }%
\newcommand{\apj}{Astroph.J. }%
\newcommand{\apjl}{Astroph.J.Lett. }%
\newcommand{\apjs}{Astroph.J.Supp. }%
\newcommand{\ao}{Appl.~Opt. }%
\newcommand{\apss}{Astroph.J.\&Sp.Sci. }%
\newcommand{\aap}{Astron.\&Astroph. }%
\newcommand{\aapr}{Astron.\&Astroph.~Rev. }%
\newcommand{\aaps}{Astron.\&Astroph.~Suppl. }%
\newcommand{\aj}{Astron.Journ. }%
\newcommand{\azh}{AZh }%
\newcommand{\memras}{MmRAS }%
\newcommand{\mnras}{Mon.Not.Royal~Astr.~Soc. }%
\newcommand{\na}{New Astron. }%
\newcommand{\nar}{New Astron. Rev. }%
\newcommand{\pra}{Phys.~Rev.~A }%
\newcommand{\prb}{Phys.~Rev.~B }%
\newcommand{\prc}{Phys.~Rev.~C }%
\newcommand{\prd}{Phys.~Rev.~D }%
\newcommand{\pre}{Phys.~Rev.~E }%
\newcommand{\prl}{Phys.~Rev.~Lett. }%
\newcommand{\pasa}{PASA }%
\newcommand{\pasp}{Proc.Astr.Soc.Pac. }%
\newcommand{\pasj}{Proc.Astr.Soc.Jap. }%
\newcommand{\rpp}{Rep.Prog.Phys. }%
\newcommand{\skytel}{Sky\&Tel. }%
\newcommand{\solphys}{Sol.~Phys. }%
\newcommand{\sovast}{Soviet~Ast. }%
\newcommand{\ssr}{Space~Sci.~Rev. }%
\newcommand{\nat}{Nature }%
\newcommand{\iaucirc}{IAU~Circ. }%
\newcommand{\aplett}{Astrophys.~Lett. }%
\newcommand{\apspr}{Astrophys.~Space~Phys.~Res. }%
\newcommand{\nphysa}{Nucl.~Phys.~A }%
\newcommand{\physrep}{Phys.~Rep. }%
\newcommand{\procspie}{Proc.~SPIE }%
         
\newcommand\newblock{}

\bibliographystyle{aip}

\end{document}